%% file: mainfile.tex
\documentclass[conference]{IEEEtran}
\IEEEoverridecommandlockouts

\usepackage{graphicx}
\usepackage{hyperref}

\usepackage{amsmath}
\usepackage{amssymb}
\usepackage{amsfonts}

\usepackage[noadjust]{cite}
\usepackage[inline]{enumitem}

\usepackage{mathtools}
\usepackage{multicol}
\usepackage{multirow}
\usepackage{xcolor}
\usepackage{bm}
\usepackage{subcaption}

\usepackage{tikz}
\usetikzlibrary{positioning,shadows.blur,arrows,arrows.meta,automata,positioning,fit,backgrounds}
\usepackage{pgfplots} 
\pgfplotsset{compat=newest} 
\pgfplotsset{plot coordinates/math parser=false}

\usepackage{scalerel}
\usetikzlibrary{svg.path}
\definecolor{orcidlogocol}{HTML}{A6CE39}
\tikzset{
  orcidlogo/.pic={
    \fill[orcidlogocol] svg{M256,128c0,70.7-57.3,128-128,128C57.3,256,0,198.7,0,128C0,57.3,57.3,0,128,0C198.7,0,256,57.3,256,128z};
    \fill[white] svg{M86.3,186.2H70.9V79.1h15.4v48.4V186.2z}
                 svg{M108.9,79.1h41.6c39.6,0,57,28.3,57,53.6c0,27.5-21.5,53.6-56.8,53.6h-41.8V79.1z M124.3,172.4h24.5c34.9,0,42.9-26.5,42.9-39.7c0-21.5-13.7-39.7-43.7-39.7h-23.7V172.4z}
                 svg{M88.7,56.8c0,5.5-4.5,10.1-10.1,10.1c-5.6,0-10.1-4.6-10.1-10.1c0-5.6,4.5-10.1,10.1-10.1C84.2,46.7,88.7,51.3,88.7,56.8z};
  }
}
\newcommand\orcidicon[1]{\href{https://orcid.org/#1}{\mbox{\scalerel*{
\begin{tikzpicture}[yscale=-1,transform shape]
\pic{orcidlogo};
\end{tikzpicture}
}{|}}}}
\def\x{\mathbf{x}}

\def\arXivPrint{1}  % Comment out to exclude the arXiv.org title page for IEEE

\begin{document}

\title{Run-Time Monitors Design for Adaptive Radar Systems: A Practical Framework}

\author{\IEEEauthorblockN{Pepijn Cox \orcidicon{0000-0002-8220-7050}, Mario Coutino \orcidicon{0000-0003-2228-5388}, Giuseppe Papari, Ahmad Mouri Sardarabadi \orcidicon{0000-0002-7207-9665}, Laura Anitori \orcidicon{0000-0001-5113-1324}}
\IEEEauthorblockA{\textit{Radar Technology}, TNO, The Hague, The Netherlands\\ 
\{pepijn.cox, mario.coutinominguez, g.papari, millad.mouri, laura.anitori\}@tno.nl
}}

%% --------------------- arXiv.org title page for IEEE
%
%
\ifx\arXivPrint\undefined\else

\makeatletter
\twocolumn[{
\vspace{2cm}
This paper has been accepted for publication at the

\vspace{1cm}
\centerline{\textbf{\huge{2023 IEEE Radar Conference}}}

\vspace{4cm}
%978-1-7281-6813-5/20/\$31.00 \textcopyright 2023 IEEE \\
%\noindent\hspace*{0.5cm} DIO 10.1109/RADAR42522.2020.9114620

\vspace{1cm}
\textbf{Citation}\\
P.B. Cox, M.A. Coutino, Giuseppe Papari, Ahmad Mouri Sardarabadi, and Laura Anitori, ``\@title,'' in \textit{Proceedings of the 2023 IEEE Radar Conference}, pp --, San Antonio, Texas, USA, May 2023.

\vspace{1cm}
%\textbf{IEEE Xplore URL}\\
%\url{https://ieeexplore.ieee.org/document/9114620}

\definecolor{commentcolor}{gray}{0.9}
\newcommand{\commentbox}[1] {\colorbox{commentcolor}{\parbox{\linewidth}{#1}}}

\vspace{1cm}
\commentbox{
	\vspace*{0.2cm}
	\hspace*{0.2cm}More papers of Pepijn Cox can be found at\\
	\centerline{\large{\url{https://orcid.org/0000-0002-8220-7050}}}\\~\\~\\
	\hspace*{0.2cm}and of Mario Coutino at\\
	\centerline{\large{\url{https://scholar.google.com/citations?user=APLpE9cAAAAJ}}}\\~\\~\\
    \hspace*{0.2cm}Giuseppe Papari\\
    \centerline{\large{\url{https://scholar.google.com/citations?user=cVyGR_YAAAAJ}}}\\~\\~\\
    \hspace*{0.2cm}Ahmad Mouri Sardarabadi\\
    \centerline{\large{\url{https://ieeexplore.ieee.org/author/38466962500}}}\\~\\~\\
    \hspace*{0.2cm}Laura Anitori at\\
	\centerline{\large{\url{https://scholar.google.com/citations?user=TYspZe4AAAAJ}}}
	\vspace*{0.2cm}
}

\vspace{1cm}
\textcopyright 2023 IEEE. Personal use of this material is permitted. Permission from IEEE must be obtained for all other uses, in any current or future media, including reprinting/republishing this material for advertising or promotional purposes, creating new collective works, for resale or redistribution to servers or lists, or reuse of any copyrighted component of this work in other works.
}]
\clearpage
\makeatother

\fi
%
%
%% END: --------------------- arXiv.org Title page for IEEE

\maketitle

\begin{abstract}
\input{abstract}
\end{abstract}

\IEEEpeerreviewmaketitle

%% ----- Start importing the tex files
    \input{introduction}
    \input{RTV_in_nutshell}

    \input{RT_framework}
    \input{examples}
    \input{new_insights}

\input{conclusion}
%% ----- End importing the tex files

\bibliographystyle{IEEEtran}

\bibliography{bibliography.bib}

\end{document}

%% file: abstract.tex
Adaptivity in multi-function radar systems is rapidly increasing, especially when moving towards fully adaptive, cognitive radar systems. However, the large number of available system configurations makes the rigorous verification and certification process during the testing phase, deployment, and after hardware and software upgrades, challenging, if not infeasible. To alleviate the verification process, run-time verification can be applied to oversee the correct function of a system during its operation as done in applications where on-the-fly reconfiguration/adaptation is pervasive, e.g., spacecrafts and self-driving cars. Though possible, the application of run-time verification into a radar system is not straightforward, e.g., when verifying (adaptive) radar resource managers or performance measures, such as track initiation time in dynamic environments. The goal of this paper is to introduce a framework to identify, characterize, and map the various aspects necessary for implementing run-time verification for (components of) multi-function radar systems. The proposed framework can be used by radar practitioners and researchers for applying run-time-verification to adaptive, re-configurable radar systems. In addition, we discuss how run-time verification can be leveraged to gather new insights from operational data to improve functionalities in upcoming update cycles and present an example of a verifier designed using the introduced framework.

%% file: introduction.tex
\section{Introduction}

Modern radar sensor suites are safety-critical, multi-function systems performing various tasks such as surveillance, tracking, illumination, guidance, and classification under a wide range of dynamical environmental conditions. As a result, adaptive signal processing, arbitrary waveform generators, adaptive digital beamforming techniques, and future, highly dynamic networked operations are considered to be the key ingredients to push current multi-function radar systems beyond the state-of-the-art~\cite{Greco2018,Gurbuz2019}. Moreover, supporting rapid deployment of newly developed hardware and software components, throughout the decades of the systems’ life-time, is becoming an essential focus point. 

As the next generations of radar(s) (networks) is thus expected to be comprised of highly flexible systems, the real-time behavior and correct function of these systems should be guaranteed under ever more challenging conditions, e.g., highly manoeuvrable targets in contested and congested (responsive) environments. However, the available systems' flexibility leads to a large space of available configurations, making the rigorous validation and certification of their performance during testing, run-time operation, and after hardware and software upgrades, challenging, if not infeasible. In particular, the verification of the performance guarantees of an operational system after updates could be extremely difficult~\cite{Watts2002,Ward2013}.

To tackle the aforementioned tasks, techniques from \emph{run-time verification} (RTV)~\cite{Greco2018,Gurbuz2019} have been successfully applied in different domains. Examples of RTV are found in numerous NASA spacecrafts, e.g., see~\cite{Bensalem2014} and reference within. Also, Tesla employs an RTV strategy for innovation of software update cycles of self-driving capabilities~\cite{Elluswamy2020}. 

Although RTV is well known in software engineering and other related fields, the impact of RTV in the radar community is limited. Especially when using (near) real-time adaptivity for closed-loop and/or cognitive approaches. For radar systems, we foresee that RTV techniques could be utilized to monitor
\begin{enumerate*}[label={(\arabic*)}]
    \item that hardware is operating within the boundaries;
    \item that the outcome of signal processing or tracking algorithms is according to the specifications;
    \item that the statistics of the target, clutter, or interference in the environment is within expectations;
    \item or that radar resource manager complies with the specifications. 
\end{enumerate*}
When events happen that deviate from these specification, the information can directly be shown to an operator, i.e., end user of the system, to take mitigating actions and/or the information can be stored for engineers to improve/update hardware or software components in upcoming update cycles. Moreover, RTV enables rigorous and structured anomaly detection of complex systems.

Implementing RTV is not a straightforward task as its design process is not unique and the set of options available is quite large. In the literature, the presented frameworks are mostly tailored for software or protocols and/or advocate general usages of particular algorithms, see, e.g.,~\cite{Chen2007,Zhao2009,Ahrendt2012,Bartocci2013}. As a result, this work's contribution is centred around the introduction of a framework to identify, characterize, and map various aspects necessary for implementing RTV for (components of) multi-function radar systems. This framework is inspired by~\cite{Vellegas2011}, while adapted for the challenges of RTV when applied to radar systems. The aim is to provide radar engineers and researchers a founded framework to characterize run-time verifiers to ease its later implementation. The proposed framework is applied to a detection module example to illustrate the design of a run-time verifier. Also, we will discuss how RTV can be applied to various phases in the life-cycle of a radar system and how RTV techniques can support an engineers’ understanding to develop and improve the performance of the radar system.

%% file: RTV_in_nutshell.tex
% \section{Run-Time Verification in a Nutshell} \label{sec:RV in a nutshell}
\section{Run-Time Verification Background} \label{sec:RV in a nutshell}

In this section, the necessary concepts of run-time verification are discussed. For an elaborate overview on run-time verification, the reader is referred to~\cite{Bartocci2018a,Leucker2009,Falcone2021}.

As radar systems become more complex, the exposure of their inner workings as well as their infrastructure becomes more important. Not only for analyzing long-term trends, conducting analysis, or generating of operational alerts, but also for constant development of new system features for incremental capability upkeep and enhancement.

The run-time monitor, cornerstone of RTV, allows to identify, detect, and flag
\begin{enumerate*}[label={(\roman*)}]
    \item error or failure, 
    \item deviations of the designer's assumption, e.g., use-cases not considered at design time, and
    \item scheduling of preventive maintenance.
\end{enumerate*}

Generally, RTV contains three parts
\begin{enumerate*}[label={(\roman*)}]
    \item (set of) formal \emph{specification(s)}, i.e., a concrete (textual) object describing a property,
    \item physical system or a model of the system, and
    \item the \emph{algorithm} or \emph{monitor} to verify if the specification of (i) is met by the system or the model of the system in (ii).
\end{enumerate*} 

The (set of) formal specification(s) needs to be represented by formal languages in order for run-time verification techniques to monitor it. Examples of formal languages are regular expressions, \emph{mission-time linear temporal logic} (MLTL) used for RTV of satellites~\cite{Luppen2021}, or \emph{signal temporal logic} (STL) employed on RTV of autonomous driving systems~\cite{Zapridou2020}. Treatment of formal languages is outside of the scope of this paper.

The \emph{module} of the system defines the component that will be run-time verified. \emph{Monitoring} is defined as the process of collecting and exposing (the relevant portion of) real-time data of a system. Note that the monitor \emph{only} monitors, meaning that it will \emph{not} interfere with or modify the system\footnote{Monitors could potentially interfere in highly unsafe situations.}. For the monitor to provide a verdict, the available data of the system (\emph{traces}\footnote{In software engineering, the \emph{logs}, \emph{traces} or quantitative data are categorized differently, for the sake of exposition, we only use the term trace.}) will be processed and tested against the specification(s). Note that, the overhead due to extra computations and accessibility to particular traces should also be taken into account when designing the system and deciding on the type of information that needs to be observed. 

%In Section~\ref{sec:run time framework}, a general framework is given to characterize run-time monitors.

%% file: RT_framework.tex
\section{Framework to Characterize Run-Time Verification for Radar Systems}
\label{sec:run time framework}

\begin{figure*}[!t]
    \vspace{-3mm}
    \centering
    \includegraphics[width=1\textwidth, angle=0]{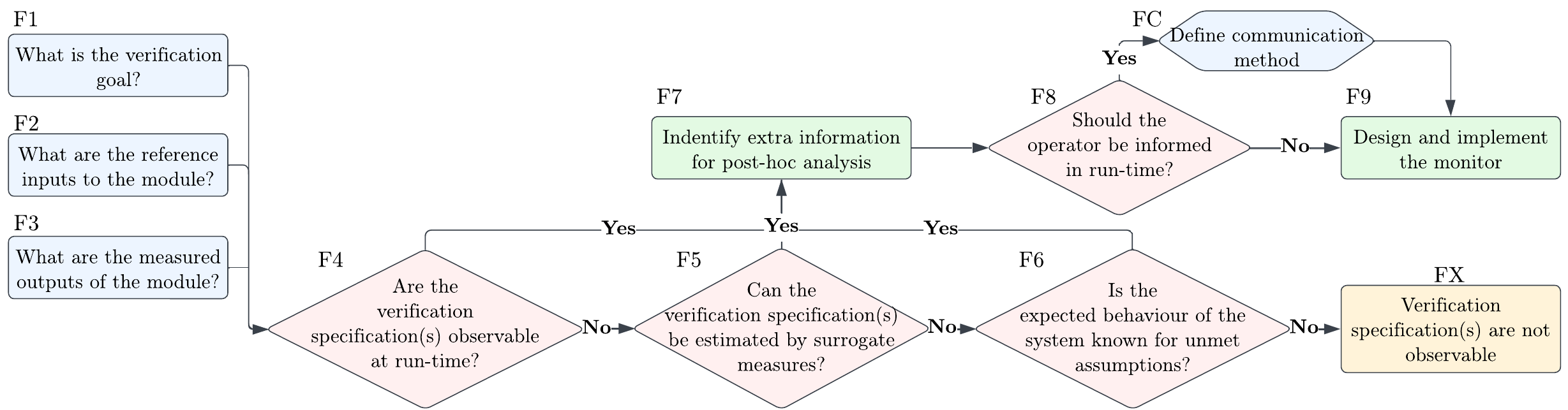}
    \caption{Illustration of the proposed framework to characterize run-time verification for adaptive radar systems.}
    \label{fig:runtime framework flow diagram}
    \vspace{-5mm}
\end{figure*}

In this section, a framework will be discussed to characterize run-time monitors with particular focus on radar systems. As previously mentioned, this framework takes inspiration from~\cite{Vellegas2011} and it is adapted to usage on radar systems. A graphical summary of the framework is shown in Fig.~\ref{fig:runtime framework flow diagram}.

As a first step, we need to identify all what is already known about the module and verification design, i.e.,
\begin{itemize}[leftmargin=*]
%
%  Monitoring goal
    \item[] [F1] \emph{\textbf{What is the run-time monitoring goal?}} The main reason or justification to implement run-time monitoring. A monitoring goal could come from the specification requirements or non-functional regulation requirements, e.g., to retain a certain performance measure, identify deviations from the design assumptions, level of reliance, security requirements, or (environmental) safety regulations, for instance, to avoid EM congestion.
%
% Reference inputs
    \item[] [F2] \emph{\textbf{What are the reference inputs to the module?}} The set of value(s) and types that characterize the information towards the module and under which specifications (bounds) on the reference inputs the module is designed. The reference inputs can consist of
        \begin{enumerate*}[label=(\roman*)]
            \item single values or signal values,
            \item statistical properties of the signals or single values,
            \item quality of service,
            \item service level agreements,
            \item constructs on defining computational states, or
            \item functional requirements.
        \end{enumerate*}
        Equally important is to identify the specifications (boundaries) and operational conditions for the reference inputs such as
        \begin{enumerate*}[label=(\alph*)]
            \item limits of a continuous domain or state,
            \item conditions for quality of service requirements or logical expressions, 
            \item external conditions or assumptions, e.g., hypothesized statistical clutter or noise properties of the reference inputs; and
            \item conditions expressing states of misbehaviour.
        \end{enumerate*}
%
% Measured outputs
    \item[] [F3] \emph{\textbf{What are the observed outputs?}} The set of values and types that can be measured in/by the system. The measurements that could be performed are
        \begin{enumerate*}[label=(\roman*)]
            \item measurements on physical properties, e.g., front-end electronics temperature;
            \item observed quality of service of the module, e.g., accuracy of the estimated target parameters; 
            \item measurements on logical properties of computational elements, e.g., number of iterations, processing time, CPU load; and
            \item measurements on external conditions, e.g., weather conditions, external object track data.
        \end{enumerate*}
\end{itemize}

\noindent After the verification goal, inputs and outputs are defined, proceed to identify the specifications (boundaries) and operational conditions under which the module is designed, i.e.,

\begin{itemize}[leftmargin=*]
%
% Verification specifications
    \item[] [F1-F3] \emph{\textbf{What are the verification specifications?}} The definition of the (set of) formal specification(s) that quantify the run-time monitoring goal under the boundary conditions on the reference inputs and measured outputs. In general, the (set of) specification(s) arises from the identified run-time verification goal.
\end{itemize}

\noindent Once the system's specification are identified, they need to be formalized by using one of the formal languages mentioned in Sec.~\ref{sec:RV in a nutshell}. Besides the common difficulties due to limitations of particular formal languages, there exist multiple ways to formalize a specification which poses an extra design challenge.

When the specifications are represented in formal languages and the involved (necessary) input and outputs identified, proceed to the design of the run-time monitor for the verification task considering:

\begin{itemize}[leftmargin=*]
%
% Measurability
    \item[] [F4] \emph{\textbf{Are the verification specification(s) observable at run-time?}} This clarifies if it is possible to verify the specifications directly from the reference inputs and measured outputs, e.g., verifying temperature with a thermal sensor.
%
% Predictability
    \item[] [F5] \emph{\textbf{Can the verification specifications be predicted or measured by surrogate measures?}} The monitor requires to estimate or predict measures by indirect means, e.g., estimating the computation time by measuring the interval of broadcasting solutions or estimating the detection performance using a simulation model under the observed weather conditions.
%
% Known beheviour with unmet assumptions
    \item[] [F6] \emph{\textbf{Is the expected behaviour of the system known for unmet specifications?}} When the verification specification(s) are not measurable nor can be estimated, then a monitor might be constructed by falsifying the specifications. For example, verifying that particular weather conditions are met might be falsified when the observed false alarm rate of the detector increases above a certain threshold.
\end{itemize}

\noindent Posterior to the characterization and design of the monitor, identify if the information collected by the monitor also allows for diagnosing and (possibly) reproducing an issue. A careful balance between too-much and too-little information should be made when collecting and compressing traces:

\begin{itemize}[leftmargin=*]
%
% Post-hoc analysis
    \item[] [F7] \emph{\textbf{Identify extra information for post-hoc analysis.}}  Characterize the set of reference input traces, observed output traces, estimated surrogate measures, or a post-processed abstraction that is rich enough in information for post-hoc module evaluation by engineers and/or operators.
%
% Operator information
    \item[] [F8] \emph{\textbf{Should the operator be informed on run-time?}} The monitor can also inform the operator on run-time when the specifications are not met. Based upon the information provided by the monitor, the operator may take mitigating actions. This step requires to consider system-wide implications, e.g., choosing the communication method [FC].
\end{itemize}

\noindent When all the previous aspects have been identified and characterized, the last step is to determine the monitoring algorithm that corresponds with the chosen formal language and the computational platform. That is,
\begin{itemize}[leftmargin=*]
%
% Defining the monitor
    \item[] [F9] \emph{\textbf{Design and implement the monitor.}} Selecting the appropriate technique or algorithm for run-time monitoring is dependent on the chosen formal language of the (set of) specifications, potential estimation procedure of the surrogate measures, and computational platform. For a detailed discussion see, e.g.,~\cite{Bartocci2018a,Tamura2013,Filieri2013,Falcone2021}.
\end{itemize}

\noindent It might happen that the flow-diagram in Fig.~\ref{fig:runtime framework flow diagram} will end in:
\begin{itemize}[leftmargin=*]
%
% No observable
    \item[] [FX] \emph{\textbf{Verification specification(s) are not observable.}} The verification goal and the corresponding metric(es) are not observable, which implies that the (set of) specification(s) and metric(es) are not measurable nor can be estimated from the traces given to the monitor provided by the system. A redesign of the system needs to be started, e.g., by making other reference inputs or measured outputs available to the monitor.
\end{itemize}

\noindent In such cases, after redesign, the presented framework [cf. Fig.~\ref{fig:runtime framework flow diagram}] can be reapplied.

%% file: examples.tex
\section{A Run-Time Verification Study-Case: Detector Performance} \label{sec:examples}

To illustrate the proposed framework, in this section, a monitor for a single-target radar detector is designed. The specification of the monitor is not directly observable and, hence, it will be predicted aiding a simple tracker model. 

\subsection{Characterizing the Monitor} \label{subsec:examp char monitor}

The characterization of the monitor for the detector module will be done based on the proposed framework in Sec.~\ref{sec:run time framework}. For this imaginary, simple, and illustrative example, consider a single-detection case where the detector either provides us with a detection or a false positive at each time step.

\noindent\emph{\textbf{Verification goal:}} The detector should not exceed a maximum level of false positives.

\noindent\emph{\textbf{Reference inputs:}} Time series of complex values that correspond to signals plus thermal noise. The time series cannot have infinite values and the noise should be circular complex Gaussian distributed.

\noindent\emph{\textbf{Measured outputs:}} Stream of positions of estimated targets. The position are bounded and cannot have an infinite value.

\noindent\emph{\textbf{Specifications:}} Due to the stochastic nature of the false positive of the detector, the specifications are defined as the \emph{probability} of a threshold-crossing event:
\begin{equation} \label{eq:DefSpec}
    Pr(f_{FP}\leq T_{FP})\geq c_1,
\end{equation}
where $Pr$ denotes the probability, $T_{FP}$ is the maximum value of the false positive rate $f_{FP}$ and $c_1$ represents the minimum confidence level. If~\eqref{eq:DefSpec} is satisfied, then the detector is operating as specified.

\noindent\emph{\textbf{Estimating surrogate measure:}} Obtaining the probabilistic measure in~\eqref{eq:DefSpec} is discussed in Sec.~\ref{subsec:estimating measure}.

\noindent\emph{\textbf{Identified extra information for post-hoc analysis:}} The monitor should store the reference inputs and measured outputs of the previous 20 time steps including the current.

\noindent\emph{\textbf{Inform the operator in run-time.}} When~\eqref{eq:DefSpec} is falsified then inform the operator either if the false positive rate $f_{FP}$ is higher than $T_{FP}$ and/or if the confidence bound $c_1$ is not met.

\subsection{Estimating a Surrogate Measure} \label{subsec:estimating measure}

In a radar system, the true target locations are unknown to the system. Therefore, the false positives rate can only be estimated from the available data. For this illustrative example, consider a single-target track case where the detector either provides us with a detection or a false positive at each time step. Tracks generated by true positives will conform to an expected degree of smoothness based on the estimated parameters of the target, while false positives will be scattered almost randomly in our region of interest. The specification in~\eqref{eq:DefSpec} is
\begin{equation} \label{eq:Integrals}
        Pr(f_{FP}\leq T_{PF}) = \int_0^{T_{PF}} p(x) dx,
\end{equation}
where $p(\cdot)$ denotes the probability density function of the false detection. The extension to a case with multiple tracks, time-varying distributions of $f_{FP}$, and/or estimating the false negative rate\footnote{The presence of temporal gaps between two detections could be interpreted as a false negative to determine the probability of false negatives $Pr(f_{FN})$.} is possible, however, it is outside the scope of this work.

Let $\x_n$ be the position of the target at time $t_n$, and let $\hat{\x}_n$ be the predicted position of $\x_n$ based on observations $\x_1, ..., \x_{n-1}$. The prediction of $\hat{\x}_n$ can be carried out using a time sequence predictor, e.g., Kalman filter, autoregressive models, etc. The distance $\delta_n=\hat{\x}_n-\x_n$ is an indication of how smooth a trajectory is: a smaller $\vert \delta_n\vert$ indicates a smoother trajectory. The smoothness is represented by the probability $z$ that a certain detection is true and it can be estimated by assuming a Gaussian model:
\begin{equation} \label{eq:Gauss}
    z_n = Pr\{\textrm{True detection}|\delta_n\} = \frac{1}{\sigma\sqrt{2\pi}} \exp\left(-\frac{\delta_n^2}{2\sigma^2}\right),
\end{equation}
where $\sigma$ is a hyperparameter to be estimated from the data and $z_n$ is the probability associated to the $n^{th}$ detection. Let $k \leq n$ be the unknown number of true detections. Regard the pair $(n,k)$ as a characterization of the \emph{state} in which the detector is. For each new detection, the system will make one of the following two transitions, depending on whether that detection ($d$) is true ($d=1$) or false ($d=0$):
\begin{equation} \label{eq:Transitions}
    \begin{tabular}{lll}
        $d=1$ & $(n,k) \rightarrow (n+1, k+1)$ &$\text{Prob:} z_{n+1}$,\\
        $d=0$ & $(n,k) \rightarrow (n+1, k)$&$\text{Prob:} 1-z_{n+1}$. \\
    \end{tabular}
\end{equation}
    
For a new detection at the next time instance, the number of total detections increases from $n$ to $n+1$, while the number of true detections increases for $k$ to $k+1$ only if the $(n+1)^{th}$ detection is true. For each detection, only its probability to be true or not is known from~\eqref{eq:Transitions}. Therefore, after $n$ detections, the system can be in any of the states $(n,k)$, with $k$ ranging from $0$ to $n$. The probability $p_{k|n}$ of being in the state $(n,k)$ can be viewed as the conditional probability of having $k$ true detections out of $n$ in total:
\begin{equation} \label{eq:TransProb}
    p_{k|n}=
    \begin{cases}
        (1-z_n)p_{0|n-1}, & k=0,\\
        z_np_{k-1|n-1} + (1-z_n)p_{k|n-1}, & 0<k<n,\\
        z_np_{n-1|n-1}, & k=n,
    \end{cases}
\end{equation}
with initial condition $p_{0|0}=1$. It is straightforward to obtain the probability distribution $p(x|n)$ at the $n^{th}$ detection:
\begin{multline}\label{eq:pdf}
    p(x|n) = \sum_{k=0}^n p_{k|n} p(x|k,n) \\= \sum_{k=0}^n p_{k|n} \frac{(1-x)^k x^{n-k}}{B(1+n-k,1+k)}, 
\end{multline}
where $B(\cdot)$ is the Beta function. By integrating $p(x|n)$ as indicated in~\eqref{eq:Integrals}, the confidence level in \eqref{eq:DefSpec} can be computed. 

Note that the presented scheme is not a simple counting scheme of false positives. We assign a probability value to each detection that gets associated to a track and we feed them into a Markov model to evaluate the probability distributions.

\begin{figure}[t!]
    \centering
    \begin{subfigure}[t!]{0.3\columnwidth}
        \centering
        \input{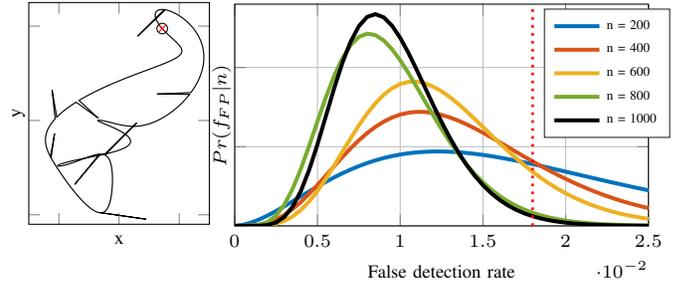}
        \vspace{0.5mm}
    \end{subfigure}\hfill%
    \begin{subfigure}[t!]{0.7\columnwidth}
        \centering
        \input{Prob.tex}
    \end{subfigure}%
    \caption{Left: a trajectory with $n=100$ detections, current position $\x_{80}$ (circle) and prediction $\hat{\x}_{80}$ (cross). Right: probability distribution of the frequency of false detections. The red dotted vertical line indicates the specification threshold.}
    \label{fig:Results}
    \vspace{-5mm}
\end{figure}

\subsection{Simulation Results} \label{subsec:simulation results}

To illustrate the discussed monitor, consider a simple example with a synthetic smooth trajectory, to which a few false detections have been added. The trajectory is a sequence of $n=1000$ consecutive detections of a target with $n_f=8$ falsely injected perturbation, resulting in a false detection rate of $n_f/n_t = 0.008$. In Fig.~\ref{fig:Results} (left), the first $80$ already observed target locations and the future $20$ observations are shown. The trajectory is mostly smooth, except for few points in which the target seems to move erratically. In Fig.~\ref{fig:Results} (right), the probability density $Pr(f_{FP}|n)$ is given for various number of detections $n$ using~\eqref{eq:pdf}. As the number of detections $n$ increases, the corresponding pdf becomes narrower around the estimated false detection rate converging to the true false detection rate of $0.008$.

The specification of the monitor in Sec.~\ref{subsec:examp char monitor} are $Pr(f_{FP}\leq T_{PF}|n)\geq c_1$ where we choose $T_{PF}=0.018$, $c_1=95\%$, and $\sigma^2$ is set as the standard deviation of the difference between the trajectory with and without perturbations. Fig.~\ref{fig:Results} shows that the specification is violated when the monitor has insufficient detection points, otherwise, the monitor accepts the specification. For this particular trace, the monitor will inform the operator that the confidence bound is not met until $n\leq800$ detections are observed.

Note that the presented monitor requires many samples to obtain the statistical confidence. For real applications, another estimator to predict $f_{FP}$ with less samples is desired. However, this is outside of the scope of this work.

%% file: Prob.tex
\definecolor{mycolor1}{rgb}{0.00000,0.44700,0.74100}%
\definecolor{mycolor2}{rgb}{0.85098,0.32549,0.09804}%
\definecolor{mycolor3}{rgb}{0.92941,0.69412,0.12549}%
\definecolor{mycolor4}{rgb}{0.46667,0.67451,0.18824}%
\begin{tikzpicture}[font=\scriptsize]

\begin{axis}[%
width=55mm,
height=29.5mm,
at={(0,0)},
scale only axis,
xmin=0,
xmax=0.025,
xlabel={False detection rate},
ymin=0,
ymax=140,
yticklabels=\empty,
ylabel={$Pr(f_{FP}|n)$},
ylabel style={yshift=-0.9em},
%axis background/.style={fill=white},
%title style={font=\bfseries},
%title={$\text{Probability density of P}_{\text{FP}}\text{ (P}_{\text{FN}}\text{)}$},
xmajorgrids,
ymajorgrids,
legend style={legend cell align=left, align=left, draw=white!15!black, font=\tiny, xshift=4mm, row sep=0pt},
%scaled ticks=false,
%tick label style={
%                /pgf/number format/fixed,
%                /pgf/number format/fixed zerofill,
%                /pgf/number format/precision=3
%            }
every x tick scale label/.style={at={(xticklabel cs:1,5pt)},yshift=0em,left,inner sep=0pt}
%every x tick scale label/.style={
%    at={(1,0)},xshift=1pt,anchor=south west,inner sep=0pt
%}
]
\addplot [color=mycolor1, line width=1.5pt]
  table[row sep=crcr]{%
0	0.0102320002908896\\
0.000500250125062531	0.712926300932798\\
0.00100050025012506	2.15096932072104\\
0.00150075037518759	4.16912789836576\\
0.00200100050025012	6.62929140698763\\
0.00250125062531266	9.4097304856367\\
0.00300150075037519	12.404190823679\\
0.00350175087543772	15.5208777992909\\
0.00400200100050025	18.6813767495589\\
0.00450225112556278	21.8195444090921\\
0.00500250125062531	24.8803993390185\\
0.00550275137568784	27.8190327556649\\
0.00600300150075038	30.5995558651049\\
0.00650325162581291	33.1940954494395\\
0.00700350175087544	35.5818458894069\\
0.00750375187593797	37.7481829218262\\
0.0080040020010005	39.6838421131089\\
0.00850425212606303	41.3841631901783\\
0.00900450225112556	42.8483999294668\\
0.00950475237618809	44.0790941962048\\
0.0100050025012506	45.0815118929881\\
0.0105052526263132	45.8631379700509\\
0.0110055027513757	46.4332272284983\\
0.0115057528764382	46.802407377139\\
0.0120060030015008	46.9823306540911\\
0.0125062531265633	46.9853702711742\\
0.0130065032516258	46.8243579615998\\
0.0135067533766883	46.5123589923426\\
0.0140070035017509	46.0624811273885\\
0.0145072536268134	45.4877141850387\\
0.0150075037518759	44.8007970118971\\
0.0155077538769385	44.014108890085\\
0.016008004002001	43.139582596312\\
0.0165082541270635	42.1886365365883\\
0.0170085042521261	41.1721235844934\\
0.0175087543771886	40.1002944509918\\
0.0180090045022511	38.9827736074295\\
0.0185092546273137	37.8285459687563\\
0.0190095047523762	36.6459527200273\\
0.0195097548774387	35.4426948350452\\
0.0200100050025013	34.2258429910469\\
0.0205102551275638	33.0018527275302\\
0.0210105052526263	31.7765838305688\\
0.0215107553776888	30.5553230465543\\
0.0220110055027514	29.3428093414436\\
0.0225112556278139	28.1432610237569\\
0.0230115057528764	26.9604041422396\\
0.023511755877939	25.7975016527268\\
0.0240120060030015	24.6573829239502\\
0.024512256128064	23.5424732193626\\
0.0250125062531266	22.4548228520302\\
};
\addlegendentry{n = 200}

\addplot [color=mycolor2, line width=1.5pt]
  table[row sep=crcr]{%
0	4.55069876066742e-08\\
0.000500250125062531	0.0226886187428079\\
0.00100050025012506	0.217266579767368\\
0.00150075037518759	0.819043033385395\\
0.00200100050025012	2.0508682238581\\
0.00250125062531266	4.07058242993447\\
0.00300150075037519	6.95113682396091\\
0.00350175087543772	10.6811547690255\\
0.00400200100050025	15.1770986598294\\
0.00450225112556278	20.3008999405179\\
0.00500250125062531	25.878980922957\\
0.00550275137568784	31.7201430959951\\
0.00600300150075038	37.6309145486387\\
0.00650325162581291	43.4277281902495\\
0.00700350175087544	48.9458206679129\\
0.00750375187593797	54.0450648384041\\
0.0080040020010005	58.6131297571422\\
0.00850425212606303	62.5664437077903\\
0.00900450225112556	65.8494504761466\\
0.00950475237618809	68.4326214412316\\
0.0100050025012506	70.3096340935598\\
0.0105052526263132	71.4940640382063\\
0.0110055027513757	72.015871080955\\
0.0115057528764382	71.9178962390388\\
0.0120060030015008	71.2525288045893\\
0.0125062531265633	70.0786526329373\\
0.0130065032516258	68.4589392176588\\
0.0135067533766883	66.457521693164\\
0.0140070035017509	64.138058056384\\
0.0145072536268134	61.5621727508384\\
0.0150075037518759	58.7882523289995\\
0.0155077538769385	55.8705622114645\\
0.016008004002001	52.8586466559127\\
0.0165082541270635	49.7969720860821\\
0.0170085042521261	46.7247741696361\\
0.0175087543771886	43.6760708462281\\
0.0180090045022511	40.6798063760586\\
0.0185092546273137	37.7600949898197\\
0.0190095047523762	34.9365365499862\\
0.0195097548774387	32.2245805367084\\
0.0200100050025013	29.6359184714485\\
0.0205102551275638	27.1788884655926\\
0.0210105052526263	24.8588788508501\\
0.0215107553776888	22.6787207694991\\
0.0220110055027514	20.639062157597\\
0.0225112556278139	18.7387177449994\\
0.0230115057528764	16.9749915378309\\
0.023511755877939	15.3439697664105\\
0.0240120060030015	13.8407835047578\\
0.024512256128064	12.4598411295424\\
0.0250125062531266	11.1950315202065\\
};
\addlegendentry{n = 400}

\addplot [color=mycolor3, line width=1.5pt]
  table[row sep=crcr]{%
0	7.46065355003266e-17\\
0.000500250125062531	0.000470946087639637\\
0.00100050025012506	0.0162273698927314\\
0.00150075037518759	0.124644376671179\\
0.00200100050025012	0.50310304996283\\
0.00250125062531266	1.4151691526351\\
0.00300150075037519	3.15647270528523\\
0.00350175087543772	5.98779407787575\\
0.00400200100050025	10.0783643479403\\
0.00450225112556278	15.4718864628545\\
0.00500250125062531	22.0781206010653\\
0.00550275137568784	29.686384148732\\
0.00600300150075038	37.9941212376019\\
0.00650325162581291	46.6430414381611\\
0.00700350175087544	55.2562626063711\\
0.00750375187593797	63.471594068521\\
0.0080040020010005	70.9679688743223\\
0.00850425212606303	77.4837139876935\\
0.00900450225112556	82.8266573486896\\
0.00950475237618809	86.8769634024118\\
0.0100050025012506	89.5840962511758\\
0.0105052526263132	90.9595044974419\\
0.0110055027513757	91.0665893553896\\
0.0115057528764382	90.0093389032206\\
0.0120060030015008	87.9207547085414\\
0.0125062531265633	84.9519145819497\\
0.0130065032516258	81.2622430002079\\
0.0135067533766883	77.0113209758406\\
0.0140070035017509	72.3523709823171\\
0.0145072536268134	67.4274030595327\\
0.0150075037518759	62.3639031253385\\
0.0155077538769385	57.2728783229044\\
0.016008004002001	52.2480399487782\\
0.0165082541270635	47.3658948125636\\
0.0170085042521261	42.6865238947263\\
0.0175087543771886	38.2548468901493\\
0.0180090045022511	34.1021977197577\\
0.0185092546273137	30.2480655303475\\
0.0190095047523762	26.7018852792286\\
0.0195097548774387	23.4647898017825\\
0.0200100050025013	20.5312601052093\\
0.0205102551275638	17.8906319117677\\
0.0210105052526263	15.528434004285\\
0.0215107553776888	13.4275478164937\\
0.0220110055027514	11.5691882617326\\
0.0225112556278139	9.93371341482273\\
0.0230115057528764	8.50127580840263\\
0.023511755877939	7.25233123379076\\
0.0240120060030015	6.16802247792017\\
0.024512256128064	5.23045576799468\\
0.0250125062531266	4.42288716714036\\
};
\addlegendentry{n = 600}

\addplot [color=mycolor4, line width=1.5pt]
  table[row sep=crcr]{%
0	9.48950037955461e-17\\
0.000500250125062531	0.00274096607449358\\
0.00100050025012506	0.0930928815293881\\
0.00150075037518759	0.674497565075792\\
0.00200100050025012	2.53099724890699\\
0.00250125062531266	6.57311059197255\\
0.00300150075037519	13.4821554526045\\
0.00350175087543772	23.4576512179588\\
0.00400200100050025	36.1459474599079\\
0.00450225112556278	50.7285309302435\\
0.00500250125062531	66.103644972103\\
0.00550275137568784	81.0918817458526\\
0.00600300150075038	94.614782901914\\
0.00650325162581291	105.819366489727\\
0.00700350175087544	114.141972446804\\
0.00750375187593797	119.318506086306\\
0.0080040020010005	121.355048664387\\
0.00850425212606303	120.474447599683\\
0.00900450225112556	117.052831150307\\
0.00950475237618809	111.556712157842\\
0.0100050025012506	104.487681442738\\
0.0105052526263132	96.3383902786492\\
0.0110055027513757	87.5609392210146\\
0.0115057528764382	78.5470132382797\\
0.0120060030015008	69.6180563652333\\
0.0125062531265633	61.0233133751368\\
0.0130065032516258	52.9435130964086\\
0.0135067533766883	45.4981742690174\\
0.0140070035017509	38.7548564054526\\
0.0145072536268134	32.739064933205\\
0.0150075037518759	27.4438929775964\\
0.0155077538769385	22.8388078389494\\
0.016008004002001	18.877253611811\\
0.0165082541270635	15.5029402219267\\
0.0170085042521261	12.6548287996307\\
0.0175087543771886	10.2709130331619\\
0.0180090045022511	8.29094650757843\\
0.0185092546273137	6.65828724533459\\
0.0190095047523762	5.32103168831958\\
0.0195097548774387	4.23259853078805\\
0.0200100050025013	3.35190377573318\\
0.0205102551275638	2.64324624866019\\
0.0210105052526263	2.0760003614462\\
0.0215107553776888	1.62419193531391\\
0.0220110055027514	1.26601434332543\\
0.0225112556278139	0.983326538212019\\
0.0230115057528764	0.761161735007002\\
0.023511755877939	0.587265431982728\\
0.0240120060030015	0.451673763174477\\
0.024512256128064	0.346337513621205\\
0.0250125062531266	0.264793122313958\\
};
\addlegendentry{n = 800}

\addplot [color=black, line width=1.5pt]
  table[row sep=crcr]{%
0	3.57254237590092e-22\\
0.000500250125062531	6.14606087157979e-05\\
0.00100050025012506	0.00696342287099071\\
0.00150075037518759	0.0996944758274207\\
0.00200100050025012	0.592483295990524\\
0.00250125062531266	2.15477150711582\\
0.00300150075037519	5.72166587738141\\
0.00350175087543772	12.2038147612551\\
0.00400200100050025	22.1465130736734\\
0.00450225112556278	35.4966259993933\\
0.00500250125062531	51.5575864768085\\
0.00550275137568784	69.121955799291\\
0.00600300150075038	86.7132845958157\\
0.00650325162581291	102.853889887585\\
0.00700350175087544	116.291064440643\\
0.00750375187593797	126.143767688206\\
0.0080040020010005	131.960988268982\\
0.00850425212606303	133.704034695952\\
0.00900450225112556	131.675942224174\\
0.00950475237618809	126.423443591045\\
0.0100050025012506	118.633471949425\\
0.0105052526263132	109.039925981686\\
0.0110055027513757	98.3497283417416\\
0.0115057528764382	87.1914621689965\\
0.0120060030015008	76.0857050130998\\
0.0125062531265633	65.4336656992844\\
0.0130065032516258	55.5196167430786\\
0.0135067533766883	46.522520770785\\
0.0140070035017509	38.5327868382968\\
0.0145072536268134	31.570937771476\\
0.0150075037518759	25.6058876676702\\
0.0155077538769385	20.5713716537104\\
0.016008004002001	16.3797618004109\\
0.0165082541270635	12.9330194659955\\
0.0170085042521261	10.1308838360327\\
0.0175087543771886	7.8766050731195\\
0.0180090045022511	6.08063077155372\\
0.0185092546273137	4.66267827536132\\
0.0190095047523762	3.55260046281094\\
0.0195097548774387	2.69040055119989\\
0.0200100050025013	2.02568809243751\\
0.0205102551275638	1.51680411552469\\
0.0210105052526263	1.12978461578626\\
0.0215107553776888	0.837281461429763\\
0.0220110055027514	0.617519313772455\\
0.0225112556278139	0.453336022764711\\
0.0230115057528764	0.331331104634278\\
0.023511755877939	0.241130949970306\\
0.0240120060030015	0.17476891294493\\
0.024512256128064	0.126172067988076\\
0.0250125062531266	0.0907430487144648\\
};
\addlegendentry{n = 1000}

\addplot [color=red, dotted, forget plot, line width=1pt]
  table[row sep=crcr]{%
0.018	0\\
0.018	140\\
};

\end{axis}
\end{tikzpicture}%

%% file: new_insights.tex
\section{Generating new insights from data}\label{sec:new insights}

\begin{figure*}[!ht]
    \centering
    \includegraphics[width=0.97\textwidth]{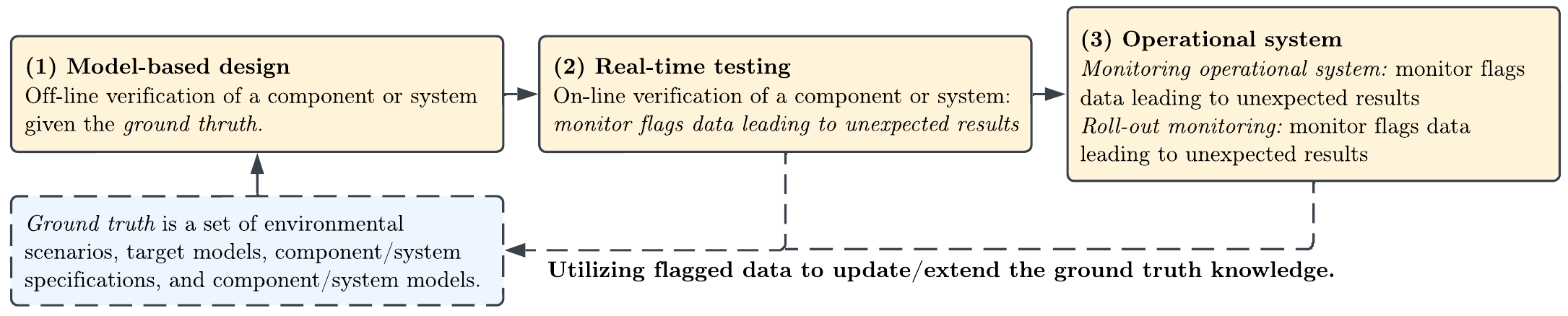}
    \caption{Run-time monitors applied in the different life-cycle phases of an adaptive, multifunction radar system.}
    \label{fig:montioring addition}
    \vspace{-5mm}
\end{figure*}

The presented run-time monitoring framework can provide benefits in multiple stages of the life cycle of the radar system. A system is designed based on a priori defined ``ground truth'', i.e., a priori defined set of
\begin{enumerate*}[label=(\alph*)]
    \item environmental scenarios, e.g., sea/land clutter condition, atmospheric propagation conditions;
    \item a set of target models, e.g., trajectories and \emph{radar cross section} (RCS) specifications;
    \item a set of component or system (performance) requirements, e.g., maximum track initiation time at $100km$ for all target models; and
    \item a set of component or system models.
\end{enumerate*}
The ``ground truth'' is used to design the hardware and software of the radar system. Defining and formalizing the ``ground truth'' is an iterative process\footnote{We could even argue that the ``ground truth'' is always an abstraction from the physical world and, therefore, it is always incomplete.} that is improved, altered, or updated based on new insights.

Before producing a radar system, the system and its sub-systems are designed via a \emph{model-based design} phase and they are designed based on a given ``ground truth'', see Fig.~\ref{fig:montioring addition}. In this phase, the majority of the activities are performed in a simulation/modelling environment potentially with some small additional proof-of-concept physical tests. During this phase, the monitors described in our framework in Sec.~\ref{sec:run time framework} are designed and implemented. Typically, systems are validated by extensive Monte Carlo testing where also the correct functioning of the monitors will be evaluated. 

In the second phase of the life-cycle, i.e., the \emph{real-time testing} in Fig.~\ref{fig:montioring addition}, the radar system including the run-time monitors is built and validated, verified, and certified in controlled environments. We foresee that monitors can be applied to \emph{flag} data series and provide an in depth analysis that led to unexpected or unintended results and, therefore, make the certification process more efficient. Additionally, as shown in Fig.~\ref{fig:montioring addition}, the flagged data series can be analysed by an engineer and it can be used to update or improve the existing set of scenarios, conditions and models or to identify new sets that have not been thought of during the design time both improving the system modules and run-time verifier. 

In the third phase of the life-cycle, the radar is delivered to the customer and it becomes \emph{operational}. For this phase, we foresee the largest potential in applying run-time verification, as monitors flag data series that lead to unexpected results in the operational phase. Storing this information is currently not a standard during operation. Having access to this information will provide a large pool of additional information about the system and its operational conditions. This flagged data can be used in an off-line process to update/extend the knowledge about the ``ground truth'' by an engineer. Extending the ``ground truth'' will, in an update cycle, be used to improve and robustify the modules in the radar. Also, the run-time monitor can be used to take immediate actions by an operator, e.g., to directly improve the radar performance by changing components or adjusting algorithmic parameters.

\subsection{Run-time Verification for Update Cycles}

Currently, large radar systems undergo update cycles only every so many years or even decades of its lifetime. With current rapid developments in both software and hardware engineering, shorter time-to-deployment is becoming essential. Hence, the radar systems should undergo regular update cycles. With shorter time between update cycles and the foreseen adaptive radar systems, real-time testing and certification for each new deployment of the system under real operational conditions becomes tedious or even impossible. Here, RTV can be an enabling technology to speed up the so-called \emph{roll-out} of new/modified modules in the operational system. The roll-out on an operational system could lower the certification costs significantly and shorten time-to-deployment.

First, note that the new/updated modules that need to be verified can be roughly categorized as \emph{active} or \emph{passive} modules. The passive modules can be verified without interfering with normal operation, e.g., verifying updated or new software modules. Verification of active modules requires interference with normal operation, e.g., when verifying radiating elements, new waveform designs. In the following, the system is updated from version $A$, i.e., its current, stable version of the radar system, to version $B$, the version that contains updated or changed modules. Note that, it is assumed that version $B$ has already passed certification tests under controlled conditions.

To address the roll-out, two deployment strategies using run-time verification are discussed that will minimize the impact to the current stable version of the deployed system. More specifically, the \emph{shadow}~\cite{shadow,Elluswamy2020} and \emph{canary} roll-out~\cite{canary} strategies will be discussed.

\noindent{\textbf{\emph{Shadow roll-out}}} will remain with $A$ as the operational version. Version $B$ is executed in the background and its outputs are only used for verification purposes and, hence, the name shadow roll-out. Shadow roll-out is only applicable to passive modules and this strategy will require higher computational loads as two versions are running in parallel.

The system remains operational as shadow deployment does not interfere with the radar operation. The usage of run-time monitors is essential to flag all unexpected results, as engineers can, most likely, only assess the information off-line.

\noindent{\textbf{\emph{Canary (staged) roll-out}}} consists on progressively deploying $B$ to a small number of users to gather traces of their run-time verification monitors to obtain system performance's feedback. The canary (staged) roll-out can be applied to update both passive and active modules.

When the system can quickly switch between $A$ and $B$, the following strategy can be used for non-critical operation. Either for a period of time or once every so many bursts, the system switches from $A$ to $B$. The run-time monitor flags potential unexpected behavior. With this strategy, active components, e.g., new digital beam tapers, new waveform designs, can be verified under a large variety of operational condition. For example, let us consider a navy with multiple ships in the same class equipped with the same type of radar systems. Some of these vessels are in service, while others are kept for training or are docked for maintenance. The non-active ships can be updated to version B and be run-time verified while in training. If $B$ is accepted then it can become an operational version. As the ships of the same class rotate in duties, $B$ can be effectively rolled-out to all ships\footnote{Potentially, if version $B$ is only a software update then it could also be (remotely) rolled-out to all other ships after acceptance on a couple of ships.}.

%% file: conclusion.tex
\section{Conclusions} \label{sec:Conclusions}

We foresee that \emph{run-time verification} (RTV) becomes a necessary methodology to alleviate the verification and certification process of adaptive radar systems which undergo frequent hardware and software updates. RTV can be applied to oversee the correct function of a system during its operation in dynamic environments. This has been successfully demonstrated in applications such as space technology and self-driving cars. In this paper, a framework to identify, characterize, and map the various aspects necessary for implementing RTV for (components of) multi-function radar systems has been introduced. It has been discussed how the proposed framework can be used by radar engineers and researchers to tackle the complex task of designing RTV techniques to adaptive, re-configurable radar systems. To illustrate the usage of the framework, we have developed a run-time verifier for a radar detector module leveraging information coming from a tracker. In addition, we have discussed how RTV can be leveraged to gather new insights from operational data to improve functionalities in upcoming update cycles.

Future work will be on bringing the presented concepts of predictive RTV to the verification and certification of radar signal processing. In particular, focusing on developing quantitative monitors to provide robustness intervals for specification compliance. In addition, combining existing RTV techniques with signal processing tools to design monitors with low computational complexity will be part of future research.

%% file: mainfile.bbl
\begin{thebibliography}{10}
\providecommand{\url}[1]{#1}
\csname url@samestyle\endcsname
\providecommand{\newblock}{\relax}
\providecommand{\bibinfo}[2]{#2}
\providecommand{\BIBentrySTDinterwordspacing}{\spaceskip=0pt\relax}
\providecommand{\BIBentryALTinterwordstretchfactor}{4}
\providecommand{\BIBentryALTinterwordspacing}{\spaceskip=\fontdimen2\font plus
\BIBentryALTinterwordstretchfactor\fontdimen3\font minus
  \fontdimen4\font\relax}
\providecommand{\BIBforeignlanguage}[2]{{%
\expandafter\ifx\csname l@#1\endcsname\relax
\typeout{** WARNING: IEEEtran.bst: No hyphenation pattern has been}%
\typeout{** loaded for the language `#1'. Using the pattern for}%
\typeout{** the default language instead.}%
\else
\language=\csname l@#1\endcsname
\fi
#2}}
\providecommand{\BIBdecl}{\relax}
\BIBdecl

\bibitem{Greco2018}
M.~S. Greco, F.~Gini, P.~Stinco, and K.~Bell, ``Cognitive radars: On the road
  to reality: progress thus far and possibilities for the future,'' \emph{IEEE
  Signal Processing Magazine}, vol.~35, no.~4, pp. 112--125, 2018.

\bibitem{Gurbuz2019}
S.~Z. Gurbuz, H.~D. Griffiths, A.~Charlish, M.~Rangaswamy, M.~S. Greco, and
  K.~Bell, ``An overview of cognitive radar: past, present, and future,''
  \emph{IEEE Aerospace and Electronic Systems Magazine}, vol.~34, no.~12, pp.
  6--18, 2019.

\bibitem{Watts2002}
S.~Watts, H.~D. Griffiths, J.~R. Holloway, A.~M. Kinghorn, D.~G. Money, D.~J.
  Price, A.~M. Whitehead, A.~R. Moore, M.~A. Wood, and D.~J. Bannister, ``The
  specification and measurement of radar performance,'' in \emph{Proc. of the
  int. Radar conf.}, Edinburgh, UK, Oct. 2002, pp. 542--546.

\bibitem{Ward2013}
K.~Ward, R.~Tough, and S.~Watt, \emph{Sea clutter: scattering, the {K}
  distribution and radar performance}, 2nd~ed.\hskip 1em plus 0.5em minus
  0.4em\relax IET, 2013.

\bibitem{Bensalem2014}
S.~Bensalem, K.~Havelund, and A.~Orlandini, ``Verification and validation meet
  planning and scheduling,'' \emph{Int. J. Software Tools for Technology
  Transfer}, vol.~16, pp. 1--12, 2014.

\bibitem{Elluswamy2020}
A.~Elluswamy, M.~Bauch, C.~Paye, A.~Karpathy, and J.~Polin, ``Generating ground
  truth for machine learning from time series elements,'' U.S. Patent US
  2020/0\,250\,473 A1, 2020.

\bibitem{Chen2007}
F.~Chen and G.~Ro{\c{s}}u, ``{MOP}: an efficient and generic runtime
  verification framework,'' in \emph{Proc. of the 22nd {ACM} {SIGPLAN} conf. on
  Object oriented programming systems and applications}, Montreal, Canada, Oct.
  2007, pp. 569--588.

\bibitem{Zhao2009}
Y.~Zhao and F.~Rammig, ``Model-based runtime verification framework,''
  \emph{Electronic Notes in Theoretical Computer Science}, vol. 253, no.~1, pp.
  179--193, 2009.

\bibitem{Ahrendt2012}
W.~Ahrendt, G.~J. Pace, and G.~Schneider, ``A unified approach for static and
  runtime verification: framework and applications,'' in \emph{Leveraging
  Applications of Formal Methods, Verification and Validation. Technologies for
  Mastering Change}.\hskip 1em plus 0.5em minus 0.4em\relax Heraklion, Greece:
  Springer, Oct. 2012, pp. 312--326.

\bibitem{Bartocci2013}
E.~Bartocci, R.~Grosu, A.~Karmarkar, S.~A. Smolka, S.~D. Stoller, E.~Zadok, and
  J.~Seyster, ``Adaptive runtime verification,'' in \emph{Proc. of the 3th Int.
  Conf. on Runtime Verification}, Istanbul, Turkey, Sep. 2013, pp. 168--182.

\bibitem{Vellegas2011}
N.~M. Villegas, H.~A. M\"{u}ller, G.~Tamura, L.~Duchien, and R.~Casallas, ``A
  framework for evaluating quality-driven self-adaptive software systems,'' in
  \emph{Proc. of the 6th Int. Symp. on Software Engineering for Adaptive and
  Self-Managing Systems}, Honolulu, HI, USA, May 2011, pp. 80--89.

\bibitem{Bartocci2018a}
E.~Bartocci and Y.~Falcone, Eds., \emph{Lectures on runtime
  verification}.\hskip 1em plus 0.5em minus 0.4em\relax Springer, 2018.

\bibitem{Leucker2009}
M.~Leucker and C.~Schallhart, ``A brief account of runtime verification,''
  \emph{J. of Logic and Algebraic Programming}, vol.~78, no.~5, pp. 293--303,
  2009.

\bibitem{Falcone2021}
Y.~Falcone, S.~Krsti\'{c}, G.~Reger, and D.~Traytel, ``A taxonomy for
  classifying runtime verification tools,'' \emph{Int. J. on Software Tools for
  Technology Transfer}, vol.~23, pp. 255--284, 2021.

\bibitem{Luppen2021}
Z.~A. Luppen, D.~Y. Lee, and K.~Y. Rozier, ``A case study in formal
  specification and runtime verification of a cubesat communications system,''
  in \emph{Proc. of the AIAA Scitech Forum}, Reston, VA, USA, Jan. 2021, pp.
  1--21.

\bibitem{Zapridou2020}
E.~Zapridou, E.~Bartocci, and P.~Katsaros, ``Runtime verification of autonomous
  driving systems in {CARLA},'' in \emph{Lecture Notes in Computer Science},
  2020, pp. 172--183.

\bibitem{Tamura2013}
G.~Tamura, N.~M. Villegas, H.~A. M\"{u}ller, J.~P. Sousa, B.~Becker, G.~Karsai,
  S.~Mankovskii, M.~Pezz\`{e}, W.~Sch\"{a}fer, L.~Tahvildari, and K.~Wong,
  ``Towards practical runtime verification and validation of self-adaptive
  software systems,'' in \emph{Software Engineering for Self-Adaptive Systems
  II}.\hskip 1em plus 0.5em minus 0.4em\relax Springer, 2013, pp. 108--132.

\bibitem{Filieri2013}
A.~Filieri and G.~Tamburrelli, ``Probabilistic verification at runtime for
  self-adaptive systems,'' in \emph{Assurances for Self-Adaptive
  Systems}.\hskip 1em plus 0.5em minus 0.4em\relax Springer, 2013, pp. 30--59.

\bibitem{shadow}
\BIBentryALTinterwordspacing
Microsoft, ``Shadow testing,'' Apr 2022. [Online]. Available:
  \url{https://microsoft.github.io/code-with-engineering-playbook/automated-testing/shadow-testing/}
\BIBentrySTDinterwordspacing

\bibitem{canary}
\BIBentryALTinterwordspacing
A.~Warner, S.~Davidovičwith, A.~Hidalgo, and B.~Beyer, ``Canarying releases.''
  [Online]. Available: \url{https://sre.google/workbook/canarying-releases/}
\BIBentrySTDinterwordspacing

\end{thebibliography}
